\documentclass[%
 aip,
jap,
reprint, 
floatfix,
preprintnumbers
]{revtex4-2}
\usepackage{graphicx}
\usepackage{mathptmx}
\usepackage{etoolbox}

\makeatletter
\def\@email#1#2{%
 \endgroup
 \patchcmd{\titleblock@produce}
  {\frontmatter@RRAPformat}
  {\frontmatter@RRAPformat{\produce@RRAP{*#1\href{mailto:#2}{#2}}}\frontmatter@RRAPformat}
  {}{}
}%

\def\maketitle{
\@author@finish
\title@column\titleblock@produce
\suppressfloats[t]}
\makeatother
\usepackage[english]{babel}

\newlength{\figsize}
\newlength{\subfigsize}
\begin{document}
\title{Optical Kerr nonlinearity of dielectric nanohole array metasurfaces with different hole shapes near the anapole state}

\author{Andrey V. Panov}
\affiliation{
Institute of Automation and Control Processes,
Far Eastern Branch of Russian Academy of Sciences,
5, Radio st., Vladivostok, 690041, Russia}
\email{panov@iacp.dvo.ru}
\preprint{J. Appl. Phys. \textbf{134}, 203104 (2023) \doi{10.1063/5.0170262}}

\date[Submitted: ]{2 August 2023}%
\accepted[Accepted: ]{5 November 2023}%
\published[Published Online: ]{27 November 2023}

\begin{abstract}
At present,
optical anapole resonances in nanostructures have attracted increasing attention due to the strong field confinement and substantially suppressed scattering. 
This study provides the results of three-dimensional finite-difference time-domain simulations exhibiting the possibility of the anapole state in high refractive index dielectric nanohole array metasurfaces having different profiles of the holes 
(square, hexagon, and octagon). 
Behavior of the effective optical Kerr nonlinearity of the metasurfaces in the vicinity of the anapole state is investigated.
Depending on the geometry, the absolute value of the effective nonlinear Kerr coefficient of the metasurface may be up to three orders of magnitude greater than that of the unstructured film.
A square transverse section of the nanohole is preferable for the optical Kerr effect in the holey metasurfaces.
The effect of the random rotation of the square holes representing the metasurface irregularity on the optical nonlinearity is examined.
As a result, the dielectric nanohole array metasurfaces display a concrete possibility to have the anapole state with large enhancement of the optical nonlinearity.
\end{abstract}

\maketitle 

\section{Introduction\label{lattcubhol_int}}
During 
the past few years, the optical effects arising from toroidal electrodynamics in nanostructures have been actively investigated by researchers. 
Theoretically, the toroidal multipoles used in analysis of the  toroidal electrodynamics in nanophotonics are combinations of the higher order terms of an expansion of the multipolar coefficients of electric parity with respect to the electromagnetic size of the source \cite{Fernandez17}.
However, the toroidal multipoles stay a useful tool for studying the optical properties of the nanoobjects.

As a rule, the electric toroidal dipole being the lowest toroidal mode is widely exploited in nanophotonics.
The electric toroidal dipole moment is associated with the poloidal currents flowing on the surface of a torus along its meridians.
The pattern of the scattering from toroidal multipole moments may coincide with one from corresponding electric or magnetic multipole moments.
Typically, the electric dipole and  electric toroidal dipole moments are used.
With specific geometric and material parameters, the scattered fields from the object of both moments are in antiphase.
This leads to destructive interference of the scattered fields, which are strongly suppressed.
Under this condition, the optical anapole state is observed.
Frequently, the anapole state is accompanied by strongly enhanced electromagnetic fields inside the nanostructure.
Such enlargement of the field intensities by orders of magnitude is essential for nonlinear optics.
Additional information about anapoles or more generally nonradiating states, including bound states in continuum, can be found in reviews \cite{Koshelev19,Savinov19,Saadabad22}.

In most cases, for nonlinear optics the anapole modes of dielectric nanoobjects were utilized for the enhanced second- \cite{Rocco18,Timofeeva18,LiHuang20} or third-harmonic \cite{Grinblat16} conversion efficiencies by the orders of magnitude of those of unstructured materials.
This is essential for producing modern laser sources.
Besides  the harmonic conversion, the anapole modes of dielectric nanoobjects can enhance other nonlinear optical effects.
Based on field confinement in InGaAs nanodisks, anapole nanolasers were proposed~\cite{Gongora17} which have the possibility to couple light into waveguide channels with four orders of magnitude intensity than classical nanolasers.
The Raman scattering intensity by an Si disk array at the anapole state was enlarged by two orders of magnitude  compared to an unpatterned Si film \cite{Baranov18}. 
Silicon nanodisks display enhancement of anapole-mediated photothermal nonlinearity by three orders of magnitude as compared with that of bulk Si \cite{Zhang20}.
Moreover, a hybrid anapole system combining a nanohole silicon disk and a longitudinal bonding dipole plasmon mode-supported plasmonic dimer exhibited  electric field enhancement in the gap region exceeding four orders of magnitude which is crucial for strong light--matter coupling \cite{Li23}.

Another nonlinear phenomenon---the optical Kerr effect (OKE) or the intensity-dependent refractive index---is essential for designing optical limiters, ultrafast optical switch devices, Kerr-lens modelocked  femtosecond lasers \cite{Yefet13}, etc.
Due to OKE, the refractive index $n$ is proportional to the light intensity $I$,
\begin{equation}
 n=n_0+n_2 I,
\label{kerrdef}
 \end{equation} 
where $n_0$ is the linear refractive index and $n_2$ is the second-order nonlinear refractive index \cite{Boyd03}.
In addition to the above-mentioned well-known applications of OKE, it was suggested for nonreciprocal optical metasurfaces, which are used for an optical diode for free-space optical signals \cite{Lawrence18} or asymmetric reflections in forward and backward light propagations caused by space--time phase modulation \cite{Guo19}. Silicon chiral metasurfaces with Kerr nonlinearities were indicated to facilitate polarization-state modulators with simple planar structures \cite{Kang23}. 

The anapole excitation of crystalline gallium phosphide (GaP) nanodisks allowed Grinblat et al. to achieve efficient ultrafast all-optical modulation due to the optical Kerr effect and the two-photon absorption in the visible and near-infrared ranges having maximum modulation depths of up to 40\%~\cite{Grinblat20}.
As shown by numeric modeling in Ref.~\cite{Panov20}, in the vicinity of the anapole state, the effective optical Kerr nonlinearity of nanodisk arrays increases by orders of magnitude.

Initially, the optical anapoles were observed for standalone nanoobjects fabricated on a substrate \cite{Miroshnichenko15,Grinblat16}.
Such technology is not very useful for the transmission optics.
Nowadays, freestanding dielectric metasurfaces (membranes) can be elaborated for visible and near-infrared ranges \cite{Karvounis18,Lim21}.
These membranes can be designed as hole arrays.
For example, freestanding lithium niobate metasurfaces consisting of a square lattice of circular air holes experimentally demonstrated high second harmonic generation efficiency in the near-infrared range \cite{Qu22}.
Primarily for perforated all-dielectric anapole metamaterials, grouped nanohole arrays were proposed \cite{Ospanova18a}.
In the geometry of Ref.~\cite{Ospanova18a} the displacement currents around the holes create a toroidal electric mode, which along with a dipole electric mode, generates the anapole state.
Nevertheless, the hole arrangement of Ref.~\cite{Ospanova18a} did not show the valuable field enhancement \cite{Panov22}, which is required for nonlinear effects.

Later, homogeneous arrays of circular nanoholes inside high-index all-dielectric plates were shown by numeric simulations to have the anapole states with the confinement of the electric field and the enhancement of effective optical Kerr nonlinearity by two orders of magnitude \cite{Panov22}.
Due to perfectness of the nanohole arrays modeled in Ref.~\cite{Panov22}, it is unclear how varying the profiles of the holes in arrays or their imperfectness affect the possibility of the anapole state and the enhancement of the optical nonlinearity.
The goal of this research is to fill the gap.

\section{Simulation details}

A numerical procedure for retrieving the effective Kerr nonlinearity of nanocomposites was introduced in Ref.~\cite{Panov18}.
This technique utilizes the results of the three-dimensional finite-difference time-domain (FDTD) simulations of the Gaussian beam propagation through a sample having optical nonlinearity.
The optical phase of the transmitted beam is calculated and its change for different light intensities $I$ permits one to evaluate the nonlinear refractive index arising from OKE and defined by Eq.~\ref{kerrdef}.
It is possible to calculate $n_2$ in several points in the transmitted beam.
This makes it feasible to estimate the mean value and the standard deviation of $n_2$.
In this work, the effective nonlinear refractive index is estimated, i.e. the  nonlinear refractive index of the nanostructure.
Details of the FDTD simulations are described in the supplementary material.

Typically, the multipole analysis of the light scattering by nanoobjects in nanophotonics is performed within the long-wavelength approximation.
In fact, the sizes of the investigated objects do not satisfy the long-wavelength approximation.
Thus, in this study the electric toroidal dipole moment $\mathbf{T}$ is described by the intensity
\begin{equation}
C^\mathrm{T}=
\frac{k^4}{6\pi\varepsilon_0^2 |\mathbf{E}_{\mathrm{in}}|^2}\left|
\mathbf{T}
\right|^2
,\quad
\label{tormoment_exact}
\mathbf{T}=
\int d\mathbf{r}\left\lbrace 3(\mathbf{r}\cdot\mathbf{J})
\mathbf{r}
-r^2 
\mathbf{J}
\right\rbrace\frac{j_2(kr)}{2r^2}
, 
\end{equation}
where $|\mathbf{E}_{\mathrm{in}}|$ is the electric field amplitude of the incident wave, $k$ is the wavenumber in vacuum, 
$\varepsilon_0$ is the vacuum permittivity, $j_2(kr)$ is the spherical Bessel function, and $\mathbf{J}$ is the induced electric current density. 
Formula \ref{tormoment_exact} is taken from the exact multipole analysis in Ref.~\cite{Alaee18}.
It is worth noting that $C^\mathrm{T}$ does not explicitly contribute to the total scattering cross section $C_\mathrm{sca}^\mathrm{tot}$.
In practice, the long-wavelength 
analysis
shifts the maximum of $C^\mathrm{T}$ to shorter wavelengths as compared with the exact one when the anapole state is observed (see the supplementary material).

In this investigation, the FDTD simulations are done for two wavelengths: $\lambda=532$~nm in the visible range for GaP with refractive index $n_{0\,\mathrm{in}}=3.49$ and a second-order nonlinear refractive index for GaP $n_{2\,\mathrm{bulk}}=6.5\times10^{-17}$~m$^2$/W and $\lambda=1034$~nm in the near-infrared range for Si with $n_{0\,\mathrm{in}}=3.56$, $n_{2\,\mathrm{bulk}}\approx4\times10^{-17}$~m$^2$/W. The origins of the optical parameters are given in the supplementary material.


Figure~\ref{poligonal_pores} shows schematic representations of the metasurfaces with the arrays of square, hexagonal, or octagonal nanoholes.
Here, $b_4$, $b_6$, or $b_8$ are the sides of the polygonal nanohole, $a$ is the lattice constant, and $h$ is the thickness of the metasurface.
The simulated linearly polarized along the $x$-axis Gaussian beam falls perpendicularly on the metasurface. 
The nanostructures in the modeling are surrounded by vacuum.

\begin{figure}[tbh!]
{\centering
\includegraphics[width=\subfigsize]{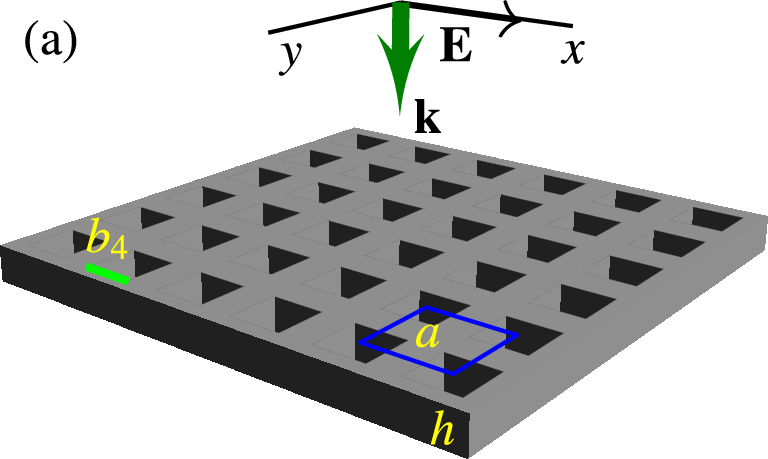}\hspace{1em}\includegraphics[width=\subfigsize]{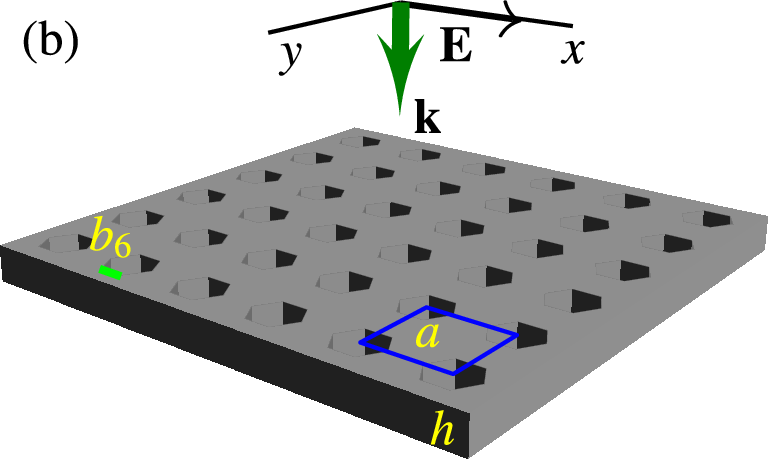}\par} 
{\centering
\includegraphics[width=\subfigsize]{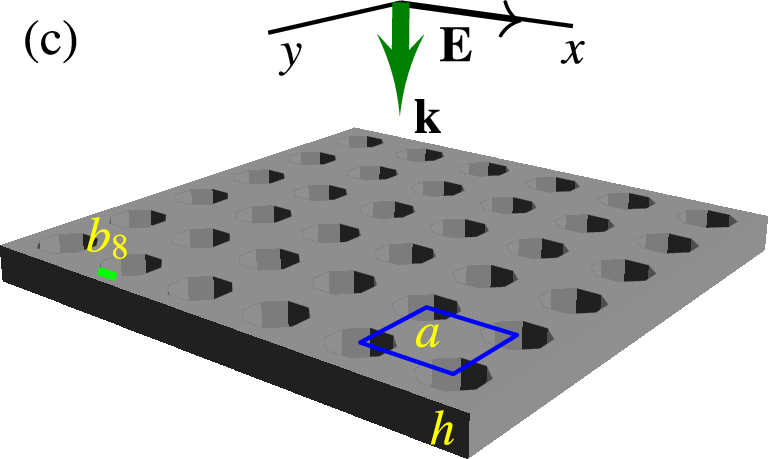}\hspace{1em}\includegraphics[width=\subfigsize]{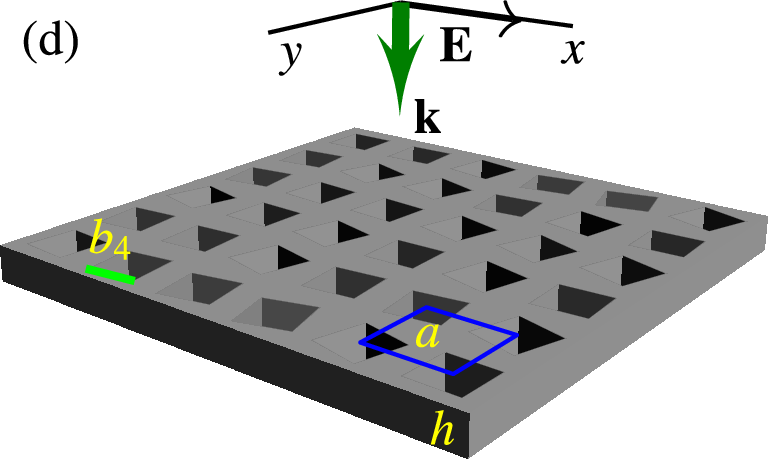}\par} 
\caption{\label{poligonal_pores}  
Schematics of the simulated freestanding metasurfaces comprising a lattice arrays of square (a), hexagonal (b), octagonal (c), or randomly rotated (d) nanoholes in a high refractive index slab. In subfigure (d), nanoholes are rotated  around their long axes with a maximum angle of rotation of $20^\circ$.
The Gaussian beam is incident normally on the metasurface. }
\end{figure} 

Before, it was shown that the GaP circular nanohole array metasurfaces possess the anapole states at $\lambda=532$~nm in some range of thicknesses \cite{Panov22}.
The anapole state in the metasurface is accompanied with large enhancement of $n_{2\mathrm{eff}}$ up to two or three orders of magnitude than that of the bulk material.
The high values of the enhancement of $n_{2\mathrm{eff}}$ for the arrays circular nanoholes in GaP were obtained for $h=100$~nm.
Analogously, the Si circular nanohole array metasurfaces at $\lambda=1034$~nm display great enhancement of $n_{2\mathrm{eff}}$ for $h=200$~nm.
For the present work, these values of the thicknesses are utilized.

\section{Results and discussion}

\begin{figure}[tbh!]
{\centering
\includegraphics[width=\figsize]{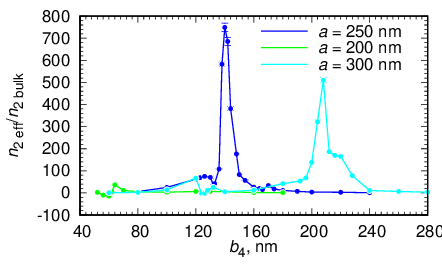}
\par} 
\caption{\label{n2_b_a_GaP_lattcubhol}
Enhancement of the effective second-order refractive index of the GaP square nanohole arrays with $h=100$~nm at $\lambda=532$~nm as functions of the nanohole side $b_4$ and the lattice parameter $a$.
}
\end{figure} 

As a result of the conducted FDTD simulations,
the dependence of the effective second-order refractive index of the GaP square nanohole arrays on the lattice parameter $a$ and the hole side size $b_4$ is illustrated by Fig.~\ref{n2_b_a_GaP_lattcubhol}.
The effective second-order refractive index of the metasurface is compared with the bulk $n_2$ of gallium phosphide.
The lattice with $a=200$~nm shows moderate enhancement of $n_{2\mathrm{eff}}$ near $b_4\approx60$~nm where the anapole mode is seen.
The highest $n_{2\mathrm{eff}}/n_{2\mathrm{bulk}}$ is observed for $a=250$~nm near $b_4\approx140$~nm.
The time-averaged electric $|\mathbf{E}|^2$ and magnetic $|\mathbf{H}|^2$ energy distributions in the transverse section of the metasurface at $h/2$ for the array of square nanoholes in GaP with $a=250$~nm, $h=100$~nm, $b_4=140$~nm are illustrated in Fig.~S1 of the supplementary material. 
These energy distributions are similar to ones obtained for the circular nanopores at the anapole state~\cite{Panov22}.
The hot spots near the corners of the squares are more pronounced than for the circles.
Figure~S2 in the supplementary material displays the electric field distribution inside the lattice element at the anapole state.
The resonances of interest in Fig.~S1 of the supplementary material
are more or less localized within the lattice elements delineated in Fig.~\ref{poligonal_pores}.
Therefore, multipole analysis for these lattice elements is utilized.

\begin{figure}[tbh!]
{\centering
\includegraphics[width=\figsize]{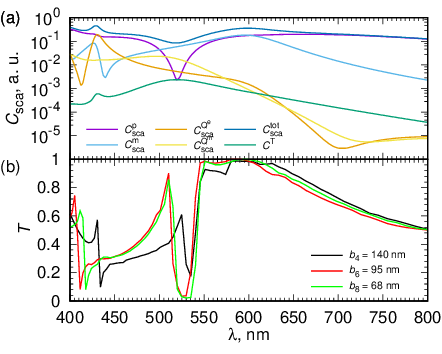}
\par} 
\caption{\label{Tspectr_GaP_lattpolyhol}
(a)
Scattering cross section spectra for the multipole contributions (electric dipole $C_\mathrm{sca}^\mathrm{p}$, magnetic dipole $C_\mathrm{sca}^\mathrm{m}$, electric quadrupole $C_\mathrm{sca}^\mathrm{Q^e}$, and magnetic quadrupole $C_\mathrm{sca}^\mathrm{Q^m}$), their sum $C_\mathrm{sca}^\mathrm{tot}$ and the intensity of the electric toroidal dipole moment $C^\mathrm{T}$ for the element
of the 
lattice with square 
nanopores in a GaP slab with 
$b_4=140$~nm.
(b)
Transmission spectra for the entire metasurfaces with polygonal nanoholes in GaP: square $b_4=140$~nm, hexagonal $b_6=95$~nm, and octagonal  $b_8=68$~nm  in the vicinity of the anapole state. 
For both graphs, refractive index $n_{0\,\mathrm{in}}$ is assumed to be constant over the whole wavelength range, $a=250$~nm and $h=100$~nm..
}
\end{figure}

The multipole decomposition of scattering cross sections for a lattice element for square nanopores at these geometric parameters is presented in Fig.~\ref{Tspectr_GaP_lattpolyhol}.
As can be seen from the figure, the intensity of the electric toroidal dipole moment $C^\mathrm{T}$ has a maximum in the scattering cross section spectrum near the wavelength of interest 532~nm at these sizes. 
At the same time, the total scattering cross section $C_\mathrm{sca}^\mathrm{tot}$ and the electric dipole cross section $C_\mathrm{sca}^\mathrm{p}$ have minima at this wavelength.
The overall transmission spectrum for the array of square nanoholes in GaP can be seen from Fig.~\ref{Tspectr_GaP_lattpolyhol}.
Near the anapole state, there is a dip in the transmission.
Even at this minimum, the transmission is about 20\%, which is a reasonable magnitude for the nonlinear optical metasurfaces at resonances \cite{Grinblat21}.
For wavelengths above, the transmission abruptly rises to the high values. 
This behavior was observed before for the anapole states in the arrays of the circular nanoholes~\cite{Ospanova18a,Panov22}.
The multipole analysis using long-wave approximation (see the supplementary material) shows that the phases of the electric and electric toroidal dipoles coincide at $\lambda=532$~nm (Fig. S5 in the supplementary material) confirming the existence of the anapole state.

It should be outlined that in practice, the radiation from the objects considered at the anapole state is not totally suppressed, for example, due to their finite size and the existence of the higher order multipoles.
In analogy with quasi-bound states in continuum, the term ``quasi-anapole'' for real-world implementations was proposed \cite{Koshelev19}.


\begin{figure}[tbh!]
{\centering
\includegraphics[width=\figsize]{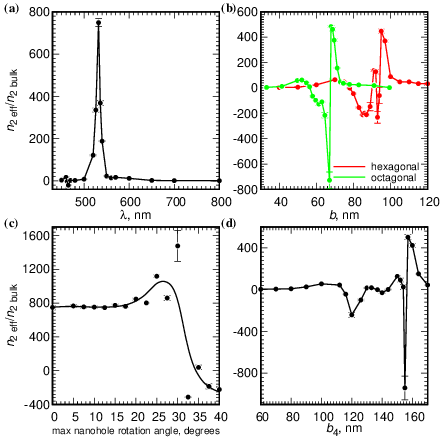}
\par} 
\caption{\label{n2_GaP_lattcubhol}
Enhancement of the effective second-order refractive index of the square (a), (c), (d) or hexa- and octagonal (b) nanohole arrays in GaP with $a=250$~nm and $h=100$~nm as functions of the wavelength (a), the maximum random rotation angle (c), and the nanohole side $b$ (b), (d). The square nanoholes are rotated randomly (c) or uniformly with the angle of $45^\circ$ (d) along their long axes. The size of the square nanopores $b_4=140$~nm in (a) and (c).
}
\end{figure} 


The spectral dependence of the enhancement of the effective OKE of the square nanohole array GaP with $a=250$~nm, $h=100$~nm, and $b_4=140$~nm is demonstrated by Fig.~\ref{n2_GaP_lattcubhol}\,(a).
In modeling, the linear refractive index $n_{0\,\mathrm{in}}$ of GaP is varied for the different values of $\lambda$ according to Ref.~\cite{Aspnes83}. 
This dependence exhibits a sharp peak at the wavelength of $532$~nm where the anapole state is observed.
In general, this curve should resemble the reciprocal relation between $n_{2\mathrm{eff}}$ and $b_4$ as the sizes and the time are linked in the FDTD simulations.


It is of interest to investigate the nonlinear optical properties of the nanopore lattices with other profile shapes.
Figure~\ref{n2_GaP_lattcubhol}\,(b)
describes the dependencies of enhancements of the effective second-order refractive index of the GaP hexagonal and octagonal nanohole arrays with $h=100$~nm at $\lambda=532$~nm on the nanohole side $b$.
These shapes show sharp dips of $n_{2\mathrm{eff}}/n_{2\mathrm{bulk}}$ for nanohole sizes just before the resonances.
Then, $n_{2\mathrm{eff}}$ steeply increases to the maximum with the growth in the size.
The anapole mode at these parameters is confirmed by the multipole decomposition (Fig.~S3), which produces the results similar to Fig.~\ref{Tspectr_GaP_lattpolyhol}
and the energy distributions (Fig.~S1).
This behavior is closer to the arrays of the circular nanoholes~\cite{Panov22} as their shapes more accurately approximate the circle.
The dip in the transmission near the anapole state is wider than that for the square nanoholes also being similar to the circular nanopores~\cite{Panov22}. 
This dip in transmission corresponds to the negative values of the effective second-order refractive index which are looked at for the arrays of the circular nanoholes~\cite{Panov22} or the disks~\cite{Panov20}.
Furthermore, a similar phenomenon was found for the second-order refractive index of the random arrangements of spheres, which inverts its sign at the Mie resonances \cite{Panov19}.
Although the metasurface with hexagonal nanopores exhibits a larger boost of the electric energy (Fig.~S1)
the enhancement of $n_{2\mathrm{eff}}$ is lower than that for the square or octagonal nanoholes.
Thus, the array of the square nanoholes in GaP shows the highest enhancement of the effective second-order refractive index with a minor dip in the transmission and without a change in the sign of $n_{2\mathrm{eff}}$ preferring this metasurface for the application in optical switches and other optical devices mentioned in Sec.~\ref{lattcubhol_int}, which are based on OKE.

It is also important to examine the stability of the nonlinear optical properties of the metasurfaces against geometric irregularities.
This seems to be more essential for the square shape of the nanoholes.
In further modeling, the square nanoholes are randomly rotated along their long axes in both directions (clockwise and counterclockwise).
The value of this rotation is limited by the maximum angle.
Figure~\ref{poligonal_pores}\,(d) describes such type of the metasurfaces.
Figure~\ref{n2_GaP_lattcubhol}\,(c)
represents the results of these simulations.
As evident from this figure,  $n_{2\mathrm{eff}}/n_{2\mathrm{bulk}}$ is stable until the maximum random angles of $20^\circ$ that is the nonlinear optical properties are prone to the geometric irregularities of the metasurface.
Such irregularities completely overlap the range of possible inaccuracies in implementing in practice the all-dielectric optical metasurfaces.
When the edges of the nanoholes have chaotic directions the effective OKE changes its sign.
This effect is coupled with the enlarged scattering and reduced transmission.


%

Another variant of the placement of the square nanohole is their rotation by the same angle of 45$^\circ$ (array of diamonds).
The dependence of the effective second-order refractive index on $b_4$ for this kind of metasurfaces is displayed in Fig.~\ref{n2_GaP_lattcubhol}\,(d).
The maximum of $n_{2\mathrm{eff}}$ is shifted to larger values of the nanohole side $b_4$. 
This dependence resembles ones for the hexagonal and octagonal nanohole arrays (Fig.~\ref{n2_GaP_lattcubhol}\,(b)).

\begin{figure}[tbh!]
{\centering
\includegraphics[width=\figsize]{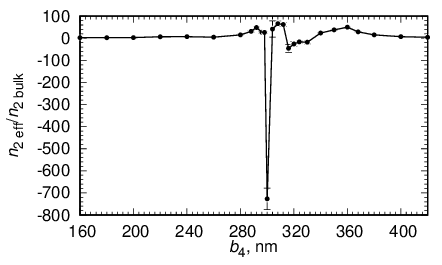}
\par} 
\caption{\label{n2_r_Si_169_lattcubhol}
Enhancement of the effective second-order refractive index of the square nanohole arrays in Si with $h=200$~nm, $a=500$~nm as a function of the nanohole side $b_4$ for $\lambda=1034$~nm.
}
\end{figure} 

\begin{figure}[tbh!]
\begin{center}
{\centering
\includegraphics[width=\figsize]{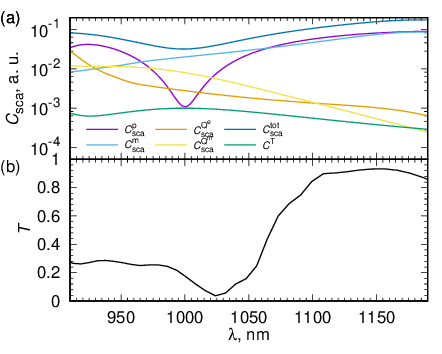}
\par
} 
\end{center}
\caption{\label{Tspectr_lattcubhol_Si}
(a)
Scattering cross section spectra for the multipole contributions (electric dipole $C_\mathrm{sca}^\mathrm{p}$, magnetic dipole $C_\mathrm{sca}^\mathrm{m}$, electric quadrupole $C_\mathrm{sca}^\mathrm{Q^e}$, and magnetic quadrupole $C_\mathrm{sca}^\mathrm{Q^m}$), their sum $C_\mathrm{sca}^\mathrm{tot}$ and the intensity of the electric dipole toroidal moment $C^\mathrm{T}$ for a lattice element of the array of the square nanoholes in Si. 
(b)
Transmission spectrum for the array of the square nanoholes in Si.
For both graphs, refractive index $n_{0\,\mathrm{in}}$ is assumed to be constant over the whole wavelength range, $a=500$~nm, $b_4=304$~nm, and $h=200$~nm.
}
\end{figure}

At the present time, silicon is frequently used for nonlinear applications in the near-infrared range.
This material may also be utilized in order to produce the  metasurfaces with nanoholes possessing the anapole state.
Figure~\ref{n2_r_Si_169_lattcubhol} represents the dependence of the effective second-order refractive index on $b_4$ for a metasurface with a lattice of the square nanopores in a silicon slab with $h=200$~nm and $a=500$~nm.
The existence of the anapole mode in the silicon metasurface with square nanoholes is confirmed by the multipole decomposition of scattering cross sections for a lattice element (Fig.~\ref{Tspectr_lattcubhol_Si}):
the intensity of the dipole electric toroidal moment $C^\mathrm{T}$ has a maximum in the scattering cross section spectrum near the wavelength of interest of 1034~nm.
It should be noted that this maximum is less prominent than that for the GaP metasurfaces in the visible range.
The transmission spectrum of the Si nanohole array with $h=200$~nm, $b_4=304$~nm and $a=500$~nm is depicted in Fig.~\ref{Tspectr_lattcubhol_Si}.
Figure~S4 in the supplementary material
illustrates the time-averaged electric $|\mathbf{E}|^2$ and magnetic $|\mathbf{H}|^2$ energy distributions in the transverse section of the lattice at $h/2$ thickness.
The distributions are similar to ones obtained above for GaP except a more pronounced additional peak of electric energy in the middle of the adjacent pores.
This peak is typical for the electric dipole moments which causes large scattering~\cite{Panov19}.
The transmission of the Si nanohole array is lower than that for the GaP metasurfaces.
Hence the optimal geometric parameters of the silicon metasurfaces with nanopores are yet to be determined before their practical implementation. 

\section{Conclusions}
To sum up, the effect of the pore shape of the nanohole lattice arrays in a high refractive index slab on the existence of the electric dipole toroidal mode is studied. 
It is demonstrated that all the investigated nanohole polygonal shapes (square, hexagonal, and octagonal) permit the possibility of the anapole state with high electromagnetic energy confinement.
The effective nonlinear Kerr coefficient for nanopore lattice arrays in proximity to the anapole state is evaluated.
The absolute value of the effective nonlinear Kerr coefficient of the metasurface is shown to be up to three orders of magnitude greater than that of the unstructured film.
It is established that the square shape of the nanohole profile provides the larger enhancement of the effective nonlinear refractive index and the better transparency providing such freestanding metasurfaces as promising light manipulating devices for transmission optics.
By the use of random rotation of the square nanoholes, it is proved that the nonlinear optical properties of the metasurface are stable with respect to its irregularities.

\section*{Supplementary material}
See the supplementary material for supporting content. The supplementary material provides a description of the geometric and optical parameters for FDTD simulations, additional graphs for the nanohole structures, and a multipole decomposition of scattering cross sections for lattice elements using formulas in long-wavelength approximation.

\begin{acknowledgments}
The results were obtained with the use of IACP FEB RAS Shared Resource Center ``Far Eastern Computing Resource'' equipment (https://www.cc.dvo.ru).
\end{acknowledgments}

\section*{Conflict of Interest}
The author has no conflicts to disclose.

\section*{Data Availability Statement}
Data underlying the results presented in this paper are not publicly available at this time but may be obtained from the author upon reasonable request.

%


\clearpage
\newpage
\setcounter{page}{1}
\setcounter{figure}{0}
\setcounter{equation}{0}
\renewcommand*{\journalname}{Supplement 1}
\renewcommand{\thefigure}{S\arabic{figure}}
\renewcommand{\thepage}{S\arabic{page}}
\title{Optical Kerr nonlinearity of dielectric nanohole array metasurfaces with different hole shapes near anapole state---Supplementary material}

\begin{abstract}
This document provides supplementary information for ``Optical Kerr nonlinearity of dielectric nanohole array metasurfaces with different hole shapes near anapole state''. 
A description of the geometric and optical parameters for FDTD simulations are given.
Additional graphs for the nanohole nanostructures are shown.
A multipole decomposition of scattering cross sections for lattice elements using formulas in long-wavelength approximation is given.
\end{abstract}

\maketitle 

\section*{FDTD simulation details}

The three-dimensional modeling of the Gaussian beam propagation through the nonlinear structure is accomplished using the Massachusetts Institute of Technology (MIT) Electromagnetic Equation Propagation (MEEP) FDTD solver \cite{OskooiRo10}.
The size of the FDTD computational domain for simulations in the visible range is $2.8\times 2.8\times 15$~$\mu$m, the space resolution of the simulations is 3.3~nm and becomes finer in the vicinity of resonances (2.5~nm). 
The size of the computational domain for simulations in the near-infrared range is $4\times 4\times 30$~$\mu$m with the resolution of  4~nm.
The modeling of light scattering by the lattice elements is performed with openEMS (Open Electromagnetic Field Solver) \cite{openEMS} and further processed with MENP (an open-source MATLAB implementation of multipole expansion for nanophotonics) \cite{Hinamoto21}.
The space resolution for the modeling with openEMS is 2.5 nm.
The linear refractive index $n_{0\,\mathrm{in}}$ for modeling  with openEMS is assumed to be constant over the whole wavelength range.

In  post-processing the results of the modeling of the light scattering, the electric and magnetic multipole scattering cross sections are calculated using MENP with the exact formulas as defined in Ref.~\cite{Alaee18}. 

Gallium phosphide is selected as material for FDTD modeling in the visible range due to its high refractive index $n_{0\,\mathrm{in}}=3.49$ at $\lambda=532$~nm with low extinction coefficient \cite{Aspnes83} which can be neglected.
The value of second-order nonlinear refractive index for GaP $n_{2\,\mathrm{bulk}}=6.5\times10^{-17}$~m$^2$/W being based on the measurements of the third-order optical susceptibility \cite{Kuhl85}.
In the near-infrared range ($\lambda=1034$~nm), silicon has similar optical properties: $n_{0\,\mathrm{in}}=3.56$, negligible absorption, $n_{2\,\mathrm{bulk}}\approx4\times10^{-17}$~m$^2$/W \cite{Bristow07}.

\newpage
\section*{Additional graphs for the GaP and Si nanostructures}

\begin{figure}[tbh!]
{\centering
{
square $b_4=140$~nm
\par}
\includegraphics[width=\subfigsize]{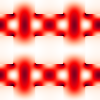}%
\includegraphics[height=\subfigsize]{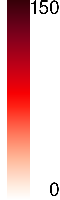}\hspace{1em}
\includegraphics[width=\subfigsize]{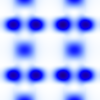}%
\includegraphics[height=\subfigsize]{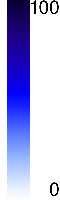}\\
{
hexagonal $b_6=95$~nm
\par}
\includegraphics[width=\subfigsize]{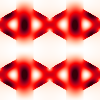}%
\includegraphics[height=\subfigsize]{colorbar_e150.eps}\hspace{1em}
\includegraphics[width=\subfigsize]{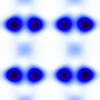}%
\includegraphics[height=\subfigsize]{colorbar_h100.eps}\\
{
octagonal $b_8=68$~nm
\par}
\includegraphics[width=\subfigsize]{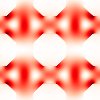}%
\includegraphics[height=\subfigsize]{colorbar_e150.eps}\hspace{1em}
\includegraphics[width=\subfigsize]{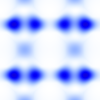}%
\includegraphics[height=\subfigsize]{colorbar_h100.eps}\\
\par
}
\caption{\label{ener_dist_GaP_361_lattpolyhol} 
Time-averaged distributions of electric $|\mathbf{E}|^2$ (left part, red color) and magnetic $|\mathbf{H}|^2$ (right part, blue color) energy densities in the arrays of polygonal nanoholes in GaP at the anapole mode ($a=250$~nm, $h=100$~nm, $\lambda=532$~nm).
The types of the nanohole transverse section and the side sizes are displayed above.
The energy densities are normalized to ones for the unperforated GaP slab.
The vertical bar shows the energy density enhancement against the latter.
The distributions are calculated within the plane at $h/2$.
The incident Gaussian beam is polarized along the vertical direction.}
\end{figure} 

\begin{figure}[tbh!]
\begin{center}
{\centering
\includegraphics[width=\figsize]{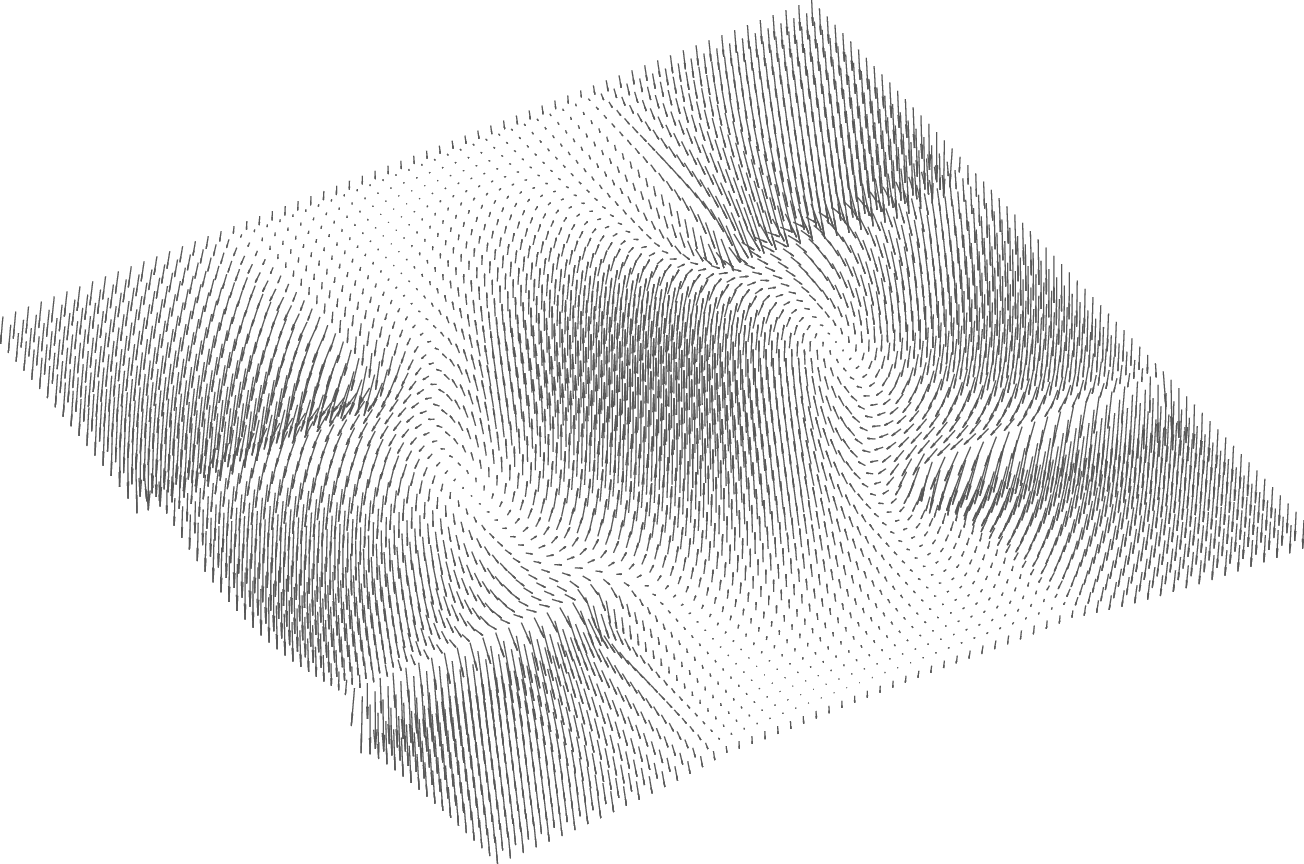}
\par
} 
\end{center}
\caption{\label{efield_dist_GaP_361_lattcubhol} 
Distribution of electric field in the element of the lattice with square nanopores array in GaP slab at the anapole mode ($a=250$~nm, $h=100$~nm, $b_4=140$~nm, $\lambda=532$~nm) within the plane of $h/2$.
}
\end{figure} 

\begin{figure}[tbh!]
\begin{center}
{\centering
{
hexagonal $b_6=95$~nm
\par}
\includegraphics[width=\figsize]{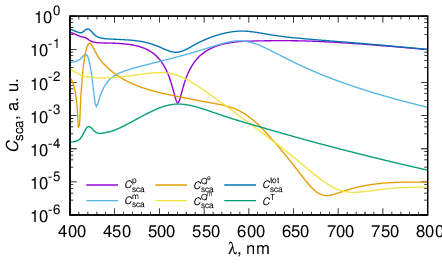}\\
{
octagonal $b_8=68$~nm
\par}
\includegraphics[width=\figsize]{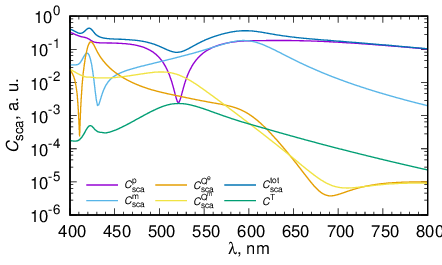}\\
\par
} 
\end{center}
\caption{\label{Csca_GaP_exact_toroidalz_log_poly_a2.5_h1}
Scattering cross section spectra for the multipole contributions (electric dipole $C_\mathrm{sca}^\mathrm{p}$, magnetic dipole $C_\mathrm{sca}^\mathrm{m}$, electric quadrupole $C_\mathrm{sca}^\mathrm{Q^e}$, magnetic quadrupole $C_\mathrm{sca}^\mathrm{Q^m}$), their sum $C_\mathrm{sca}^\mathrm{tot}$ and the intensity of the electric dipole toroidal moment $C^\mathrm{T}$ for the elements of the lattices with polygonal nanopores in GaP slab with $a=250$~nm and $h=100$~nm. 
The types of the nanohole and the side sizes are displayed above.
Refractive index $n_{0\,\mathrm{in}}$ is assumed to be constant over the whole wavelength range.
}
\end{figure} 

\begin{figure}[tbh!]
\begin{center}
{\centering
\includegraphics[width=\subfigsize]{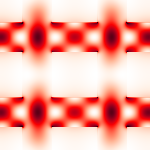}%
\includegraphics[height=\subfigsize]{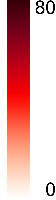}\hspace{1em}
\includegraphics[width=\subfigsize]{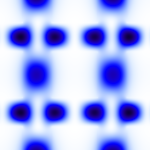}%
\includegraphics[height=\subfigsize]{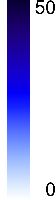}\par
} 
\end{center}
\caption{\label{ener_dist_anap_sqr_Si}
Time-averaged distributions of electric $|E|^2$ (left parts, red color) and magnetic $|H|^2$ (right parts, blue color) energy densities in the lattice of square nanoholes in Si at the anapole mode with $b_4=304$~nm, $a=500$~nm, $h=200$~nm, $\lambda=1034$~nm. 
The energy densities are normalized to ones for the unperforated Si slab.
The distributions are calculated within the plane at $h/2$.
The incident light beam is polarized along the vertical direction.}
\end{figure}

\newpage
\section*{Multipole analysis in the long-wavelength approximation}
\begin{figure}[tbh!]
\begin{center}
{\centering
{
square $b_4=140$~nm
\par}
\includegraphics[width=\figsize]{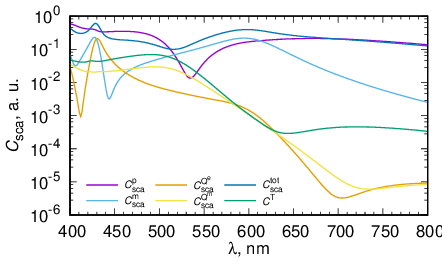}\\
{
hexagonal $b_6=95$~nm
\par}
\includegraphics[width=\figsize]{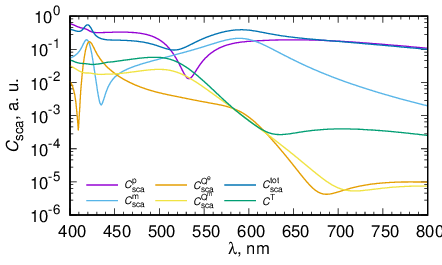}\\
{
octagonal $b_8=68$~nm
\par}
\includegraphics[width=\figsize]{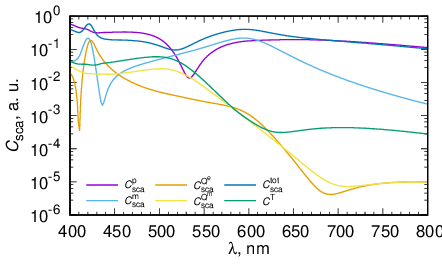}\\
\par
} 
\end{center}
\caption{\label{Csca_GaP_approx_toroidalz_log_poly_a2.5_h1}
Scattering cross section spectra for the multipole contributions (electric dipole $C_\mathrm{sca}^\mathrm{p}$, magnetic dipole $C_\mathrm{sca}^\mathrm{m}$, electric quadrupole $C_\mathrm{sca}^\mathrm{Q^e}$, magnetic quadrupole $C_\mathrm{sca}^\mathrm{Q^m}$), their sum $C_\mathrm{sca}^\mathrm{tot}$ and the intensity of the electric dipole toroidal moment $C^\mathrm{T}$ calculated using approximate equations~\cite{Alaee18} for the elements of the lattices with polygonal nanopores in GaP slab with $a=250$~nm and $h=100$~nm. 
The types of the nanohole and the side sizes are displayed above.
Refractive index $n_{0\,\mathrm{in}}$ is assumed to be constant over the whole wavelength range.
}
\end{figure} 

\begin{figure}[bth!]
\begin{center}
{\centering
\includegraphics[width=\figsize]{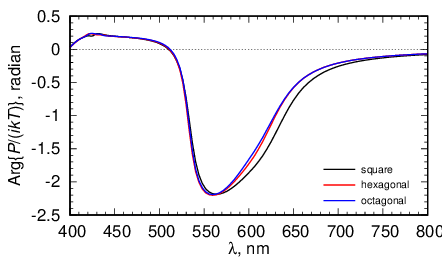}
\par
} 
\end{center}
\caption{\label{Phi_del_3shapes}
Difference of the phases between electric dipole mode $P$ and toroidal dipole mode $T$ for the lattice elements of arrays of polygonal nanoholes in GaP  with $a=250$~nm, $h=100$~nm: square $b_4=140$~nm, hexagonal $b_6=95$~nm, octagonal  $b_8=68$~nm calculated using approximate equations~\cite{Alaee18}. 
Refractive index $n_{0\,\mathrm{in}}$ is assumed to be constant over the whole wavelength range.
}
\end{figure}

In most works, a multipole decomposition of scattering cross sections in analyzing anapole states is done in long-wavelength approximation.
In many studies, the electric toroidal dipole moment $\mathbf{T}$ is described by the expression for  in the long-wavelength approximation
\begin{equation}
\label{tormoment_approx}
\mathbf{T}_\mathrm{lw}=\frac{1}{10c}
\int d\mathbf{r}\left\lbrace 3(\mathbf{r}\cdot\mathbf{J})
\mathbf{r}
-2r^2 
\mathbf{J}
\right\rbrace
, 
\end{equation}
where $\mathbf{J}$ is the induced electric current density. 
This is not quite correct as the geometrical sizes of the objects are usually larger than $\lambda/10$.
For consistency with other studies, Fig.~3 from the main article and Fig.~\ref{Csca_GaP_exact_toroidalz_log_poly_a2.5_h1} are recalculated with the use of approximate equations from Ref.~\cite{Alaee18} in Figure~\ref{Csca_GaP_approx_toroidalz_log_poly_a2.5_h1}.
As evident, the maximum of the toroidal mode in the approximated graphs is shifted to the shorter wavelengths.
Figure~\ref{Phi_del_3shapes} shows the difference of the phases between electric dipole mode $P$ and toroidal dipole mode $T$ for the lattice elements of various shapes calculated with approximate equations. 
Crossing zero in these dependencies corresponds to the anapole condition~\cite{Hinamoto21}.
In the long-wavelength approximation, this condition occurs for wavelengths just above the maximum of the electric toroidal moment. 
This condition is also shifted to the shorter wavelengths.

Moreover, currently some analysis for the infinite metasurfaces was proposed \cite{Savinov14,Cojocari21}. 
These analysis is again worked out on the top of the long-wavelength decomposition thus its application for the real objects of nanophotonics is in question.

\clearpage
%


\end{document}